# Unexpected stable stoichiometries of sodium chlorides

Weiwei Zhang[1, 2, *], Artem R. Oganov[2, 3, *], Alexander F. Goncharov[4], Qiang Zhu[2], Salah Eddine Boulfelfel[2], Andriy O. Lyakhov[2], Maddury Somayazulu[4], Vitali B. Prakapenka[5]

[1] College of Science, China Agricultural University, Beijing, 100080, P.R.China.
[2] Department of Geosciences, Department of Physics and Astronomy, State University of New York at Stony Brook, Stony Brook, NY 11794-2100, U.S.A.
[3] Geology Department, Moscow State University, 119992 Moscow, Russia.
[4] Geophysical Laboratory, Carnegie Institution of Washington, 5251 Broad Branch Road NW, Washington, D.C. 20015, U.S.A.
[5] Center for Advanced Radiation Sources, University of Chicago, Chicago, Illinois 60637, U.S.A.
*To whom correspondence should be addressed.
E-mail: zwwjennifer@gmail.com, artem.oganov@stonybrook.edu.

**At ambient pressure, sodium, chlorine, and their only known compound NaCl, have well-understood crystal structures and chemical bonding. Sodium is a nearly-free-electron metal with the bcc structure. Chlorine is a molecular crystal, consisting of $Cl_2$ molecules. Sodium chloride, due to the large electronegativity difference between Na and Cl atoms, has highly ionic chemical bonding, with stoichiometry 1:1 dictated by charge balance, and rocksalt (B1-type) crystal structure in accordance with Pauling's rules. Up to now, Na-Cl was thought to be an ultimately simple textbook system. Here, we show that under pressure the stability of compounds in the Na-Cl system changes and new materials with different stoichiometries emerge at pressure as low as 25 GPa. In addition to NaCl, our theoretical calculations predict the stability of $Na_3Cl$, $Na_2Cl$, $Na_3Cl_2$, $NaCl_3$ and $NaCl_7$ compounds with unusual bonding and electronic properties. The bandgap is closed for the majority of these materials. Guided by these predictions, we have synthesized cubic $NaCl_3$ at 55-60 GPa in the laser-heated diamond anvil cell at temperatures above 2000 K.**

**Summary:** Under pressure, sodium chloride loses its iconic simplicity, and textbook chemistry rules break down upon formation of stable compounds $Na_3Cl$, $Na_2Cl$, $Na_3Cl_2$, $NaCl_3$, and $NaCl_7$.

NaCl, the only known stable compound in the Na-Cl system, behaves regularly under pressure. However, its constituent elements show dramatic structural and electronic changes. Unlike normal materials which become metallic at sufficient compression, sodium becomes a wide-gap transparent insulator at ~200 GPa, as a consequence of orthogonality of core and valence electrons, driving localization of the latter in the interstitial regions of the structure (*1*). For chlorine, the ambient-pressure phase is orthorhombic (*Cmca*) (*2*), with theoretical calculations (Ref. 3 and this work) predicting breakdown of these molecular state metallization above ~140 GPa.

The high-pressure behavior of NaCl has been extensively studied experimentally at pressures up to 304 GPa (*4-6*) and by *ab initio* simulations (*7*), and very simple behavior was observed for NaCl - at 30 GPa the rocksalt structure was found to transform into the CsCl (B2-type) structure (*8,9*), and our calculations confirm this phase transition. Metallization of NaCl is generally expected to occur at several hundred GPa, although the existing estimates vary by several times (*10-14*): the simplest estimate based on the Herzfeld criterion (*11*) is 300 GPa, whereas DFT calculations (*14*) suggest 584 GPa and predict several ultrahigh-pressure structural transitions within the composition NaCl. Here we demonstrate stability of materials with stoichiometries other than 1:1 and complex behavior similar to that of intermetallides. This manifests the breakdown of traditional chemical rules in what was long thought to be the simplest binary chemical system.

To find the stable Na-Cl compounds and their structures, we used the *ab initio* evolutionary algorithm USPEX *(15-17)*, which has a capability to simultaneously find stable stoichiometries and the corresponding structures in multicomponent systems *(18)* (see Methods). In these calculations, any combinations of numbers of atoms in the unit cell were allowed (with the total number ≤16), and calculations were performed at pressures of 1 atm, 20 GPa, 100 GPa, 150 GPa, 200 GPa and 250 GPa. Detailed enthalpy calculations for the most stable structures allowed us to reconstruct the *P-x* phase diagram of the Na-Cl system (Fig. 1, see also Supporting Online Materials). This phase diagram features unexpected compounds - $NaCl_3$, stable above 20 GPa, $NaCl_7$, stable above 142 GPa, and $Na_3Cl_2$, $Na_2Cl$, and $Na_3Cl$, which are stable above 120 GPa, 100 GPa and 77 GPa, respectively. By stable we understand compositions on the convex hull (Fig. 1b), which have a lower free energy than any isochemical mixture of other compounds or pure elements. These new compounds are not ionic and contain no $Cl_2$ molecules. For all the newly predicted structures we computed phonons, and found them to be dynamically stable. In the entire explored pressures region, NaCl is also a stable compound, i.e. will not spontaneously decompose into other compounds. This means that to obtain the newly predicted compounds it is not sufficient just to compress NaCl, but one must do so with excess of either Na or Cl. For example, excess of $Cl_2$ is needed to enable synthesis of $NaCl_7$, e.g. through the reaction $NaCl + 3Cl_2 = NaCl_7$.

NaCl$_7$ is stable above 142 GPa and has a cubic structure (space group $Pm\bar{3}$) with 1 formula unit (f.u.) in the primitive cell. This structure is shown in Fig. 2(A) and can be described as a derivative of the A15 (β-W or Cr$_3$Si) structure type.

At 25-48 GPa, NaCl$_3$ is stable in the *Pnma* structure, which has 4 formula units in the unit cell. Unlike all the other new phases predicted here (which are metallic), this phase is a semiconductor. Its structure (Fig. 2(B)) contains almost linear asymmetric Cl$_3$ groups. Bader analysis (*19*) shows that the middle atom in the Cl$_3$-group is nearly neutral, with most negative charge on the side atoms (Table 1) and the total charge of this anion group being ~-0.8. This result is reminiscent of well-known triiodides and azides with I$_3^-$ and N$_3^-$ ions, and of recent calculations on hypothetical H$_3^-$ ions (*20*), which were found to have charge configuration [H$^{-0.81}$H$^{+0.72}$H$^{-0.81}$]$^{-0.9}$. *Pnma*-NaCl$_3$ still can be viewed as an octet compound, with covalent bonding within [Cl$_3$]$^-$-groups and ionic bonding between Na$^+$ and [Cl$_3$]$^-$. At 48 GPa, with the formation of a new phase of NaCl$_3$, the octet rule breaks down.

At 48 GPa, NaCl$_3$ is predicted to transform into a metallic A15 (Cr$_3$Si-type) structure with space group $Pm\bar{3}n$. The difference between $Pm\bar{3}n$-NaCl$_3$ and $Pm\bar{3}$-NaCl$_7$ structures is that the central atom (inside Cl-icosahedron) is Na in NaCl$_3$ (Fig. 2(C)) and Cl in NaCl$_7$ (Fig. 2(A)). Lattice parameters and bond lengths in this phase are very close (within 0.5% at 200 GPa) to those in NaCl$_7$, which implies that at this pressure Na and Cl have almost identical sizes – in stark contrast with ambient conditions, where the Cl$^-$ is much larger than Na$^+$ (ionic radii are 1.81 Å and 1.02 Å, respectively). This suggests the possibility of non-stoichiometry and disorder, with the potential for Anderson localization of electronic and phonon states.

At 200 GPa, the nearest Na-Cl distance is 2.30 Å, whereas the shortest Cl-Cl distance is 2.06 Å - only slightly longer than the bond length in the $Cl_2$ molecule (1.99 Å). However, here and in $NaCl_7$, there are no molecules, and short Cl-Cl bonds form extended monatomic chains running along the three mutually perpendicular axes. Each of these chains represents a textbook linear chain with a half-filled band – and in the free state is unstable to pairing (Peierls distortion) into $Cl_2$ molecules (*21*). The application of pressure, and influence of other chemical entities (Na and non-chain Cl atoms), stabilizes these chains in $NaCl_3$, $NaCl_7$, and elemental chlorine. Peierls theorem also explains results of our phonon calculations indicating that both $NaCl_3$ and $NaCl_7$ can exist only at high pressure and are not quenchable to atmospheric pressure and should release excess chlorine in the form of $Cl_2$ molecules.

Electronic band structures of the $NaCl_3$ and $NaCl_7$ are presented in Fig. 3(A,B), and one can see a deep and wide pseudogap for $NaCl_3$-*Pm3n*. Fig.3(C,D) shows the electron localization functions for $NaCl_7$ and $NaCl_3$. The results are unusual: in both structures Cl atoms forming the $Cl_{12}$ icosahedra show toroidal ELF maxima, corresponding to a non-closed-shell electronic configuration, while the extra Cl atom occupying the center of the icosahedron in $NaCl_7$ has a spherical ELF maximum. Thus, Cl1 and Cl2 atoms in $NaCl_7$ have different electronic structures and play very different chemical roles. Bader analysis (Table 1) confirms this and gives an unusual positive charge to the Cl1 atom located within the icosahedral cage.

The A15 structure is typical for type II high-*Tc* superconductors, which before the discovery of cuprate superconductors for decades held the record of highest known $T_C$ values and is known to be common for strong electron-phonon coupling materials.

However, our calculations for undoped A15-type NaCl$_3$ found no evidence of superconductivity.

For Na-rich side of the phase diagram, we have found several thermodynamically stable compounds – tetragonal Na$_3$Cl (space group *P*4/*mmm*, 1 f.u./cell), two phases of Na$_3$Cl$_2$: tetragonal (space group *P*4/*m*) with 2 f.u./cell and orthorhombic (space group *Cmmm*) with 2 f.u./cell, and three phases of Na$_2$Cl - tetragonal (space group *P*4/*mmm*, 2 f.u./cell), and two orthorhombic phases, *Cmmm* (4 f.u./cell) and *Imma* (4 f.u./cell).

Most of these are superstructures of the CsCl-type (B2) structure, with both Na and Cl atoms in the eightfold coordination (Fig. 4). For example, Na$_3$Cl can be represented as a [NaCl][ClCl]… sequence of layers, and the *c*-parameter of the unit cell is doubled relative to that of B2-NaCl. Na$_3$Cl$_2$ is stable above 120 GPa, and its *P4/m* structure can be described as a 1D- (rather than layered, 2D-) ordered substitution superstructure of the B2-NaCl structure.

Figure 5 shows the three phases of Na$_2$Cl predicted at different pressures; this compound shows a more complex behavior than Na$_3$Cl or Na$_3$Cl$_2$. At 100-135 GPa, the *P4/mmm* structure is stable; it is also a layered B2-type superstructure. At 135-298 GPa, Na$_2$Cl is stable in the *Cmmm* structure, and above 298 GPa in the *Imma* structure – both of which are new structure types with Na and Cl atoms in the 12- and 10-fold coordination, respectively.

The Na-Cl system gives one of the most startling illustrations of the dramatic changes of chemistry induced by pressure. In addition to the well-known insulating B2-type NaCl, semiconducting NaCl$_3$ is stable at 25-48 GPa, and metallic sodium chlorides with compositions NaCl$_7$, NaCl$_3$, Na$_3$Cl$_2$, Na$_2$Cl, Na$_3$Cl all become stable at pressures from >48 GPa (for $NaCl_3$) to >142 GPa (for $NaCl_7$) (Fig. 1). Note that these pressures are well below the predicted metallization pressure of NaCl (584 GPa).

To verify these predictions, high-pressure experiments have been performed in laser heated diamond anvil cell (DAC) at 55-60 GPa on the Na-Cl system in the excess of chlorine (see Methods). We loaded two stacked NaCl thin (5-8 μm) plates of 50 x 50 μm dimensions in the DAC cavity of 80 μm diameter and filled the rest of the cavity cryogenically by molecular chlorine. Optical absorption and Raman spectra were monitored. At 55-60 GPa due to the bandgap narrowing, $Cl_2$ became sufficiently absorptive to couple to a 1075 nm fiber laser radiation. We laser heated the NaCl plate which is insulated from the diamond anvils from all sides by chlorine, as this configuration allows heating it from both sides. Laser heating at 55-60 GPa results in a chemical reaction, which was detected by a sudden increase in temperature near and above 2000 K, which is consistent with the predicted exothermic chemical reaction. We must stress that the laser heating remains very local during this procedure as our radiometric measurements and finite element calculations show. Thus, we do not expect any reaction with a gasket material (which remains cold during the heating) or with diamond anvils; this was verified by reversibility in pressure of the Raman observations. The reaction products were examined by visual observations, optical and Raman confocal spectroscopy, and by synchrotron X-ray diffraction probes at room temperature. The results show the formation of a new material in the heated area. A new phase was found meta(stable) on pressure release down to as low as 10 GPa based on Raman spectra observations.

Visual observations of the laser heated spot shows a change in appearance as one can see a sample area with an increased reflectivity (Supplementary Fig. S6). X-ray diffraction measurements show new Bragg peaks which can be indexed in a cubic unit cell (Fig. 7). At 60 GPa the lattice parameter obtained from our XRD is a=4.6110(3) Å, which is in excellent agreement our theoretical value of 4.602 Å. The x-ray diffraction pattern also contains peaks from unreacted orthorhombic chlorine, which have been easily identified using theoretical calculation performed in this work for pure chlorine at 60 GPa. It should be noted that no trace of the B2-NaCl was found in the reaction zone and anywhere in the sample cavity, demonstrating that the chemical reaction is completed. Raman spectra of the reaction products show a set of lines distinct from those of $Cl_2$ (Fig. S7 gives a representative spectrum that shows no peaks of $Cl_2$ medium). The spectra are dependent on the observation points in intensity and even in shape. Some sample areas show narrower bands, but the overall set of lines remains the same.

These data demonstrate the formation of a new compound of Cl and Na. Based on a practically perfect agreement with theoretically predicted symmetry and lattice parameter for $NaCl_3$ and $NaCl_7$ compounds, we suggest that we synthesized a new Cl-rich cubic material, the stoichiometry of which is either $NaCl_3$ or intermediate between $NaCl_3$ and $NaCl_7$. The main Raman bands of $NaCl_3$ agreed very well with our observations, but they are broad and we see bands corresponding to all off-zone-center phonons, i.e. some of the selection rules appear to be lifted. These selection rules could be substantially lifted in surface Raman scattering (as happens in metals); another possibility to lift selection rules is to form a solid solution $(Na_{1-x}Cl_x)Cl3$ between $NaCl_3$ and $NaCl_7$. Figure S7 (Supplementary Materials) shows that the observed Raman band positions are in a good agreement with those calculated for the zone-center optical phonons of NaCl$_3$, lending further support to our conclusions. The optical absorption spectra of the synthesized material (Fig. S6 of Supplementary Materials) shows a gap-like feature at 1.7 eV, which is consistent with the predicted prominent pseudogap in the electronic density of states (Fig. 2(b)). Experiments for the Na-rich systems are underway and will be reported elsewhere.

Already in this simplest binary system, our study finds unexpected chemical compounds that the surprising low pressure of 25 GPa. Counterintuitive compounds (such as LiH$_2$, LiH$_6$, LiH$_8$) have been predicted (*22)* to appear under pressure, but experiments (*23)* failed to find them so far. Our work provides the first compelling theoretical and experimental evidence of the formation of such compounds, and violations of simple "charge balance" rule in a seemingly ionic system, already at moderate pressures. Now it should be expected that new stable compositions and new chemistry be found in the K-Cl system, and perhaps in the important planet-forming systems Mg-O, Si-O, H-C-N-O, at pressure within experimental reach or at pressures of planetary interiors.

**References and Notes:**

1. Y. Ma *et al*., *Nature* **458**, 182 (2009).
2. R. L.Collin, *Acta Crystallogr*. **5,** 431 (1952).
3. P. Li, G. Gao, Y. Ma, *J. Chem. Phys*. **137**, 064502 (2012).
4. N.Sata, G.Y.Shen, M.L. Rivers, S.R. Sutton, *Phys. Rev. B* **65,** 104114 (2002).
5. T.Sakai, E.Ohtani, N. Hirao, Y. Ohishi, *J. Appl. Phys.* **109,** 084912 (2011).
6. D.L. Heinz, R.Jeanloz, *Phys. Rev. B* **30**, 6045 (1984).
7. S. Ono, *J. Phys: Conf. Ser.* **215,** 012196 (2010).
8. S. Froyen, M.L. Cohen, *Phys. Rev. B* **29**, 3770 (1984).
9. W.A.Bassett, T.Takahash. H.K.Mao, J.S.Weaver, *J. Appl. Phys.* **39**, 319 (1968).
10. V.A.Zhdanov, V.A. Kuchin, V.V. Polyakov, *Izv Vuz Fiz.* **16**, 57 (1973).
11. M. Ross, *J. Chem. Phys.* **56**, 4651 (1972).
12. S.V. Karpenko, A.I.Temrokov, *High. Temp.* **44**, 41 (2006).
13. J. L. Feldman, B.M. Klein, M.J. Mehl, H. Krakauer, *Phys. Rev. B* **42**, 2752 (1990).
14. X. Chen, Y. Ma, *EPL* **100**, 26005 (2012)).

15. A.R.Oganov, C.W. Glass, *J. Chem. Phys.* **124,** 244704 (2006).
16. C.W.Glass, A.R.Oganov, N. Hansen, *Comput. Phys. Commun.* **175**, 713 (2006).
17. A.R.Oganov, A.O.Lyakhov, M.Valle, *Acc. Chem. Res.* **44**, 227 (2011).
18. A.R.Oganov, Y.Ma, A.O.Lyakhov, M.Valle, C.Gatti, *Rev. Mineral Geochem.* **71**, 271 (2010)
19. R.F.W.Bader, *Atoms in Molecules. A Quantum Theory.* (Oxford University Press, Oxford, 1990).
20. S.J.Grabowski, R.Hoffmann, *Chem. Phys. Chem.* **13**, 2286 (2012).
21. R.Hoffmann, *Solids and Surfaces: A Chemist's View on Bonding in Extended Structures* (VCH Publishers, New York, 1988)
22. E.Zurek, R.Hoffmann, N.W.Ashcroft, A.R.Oganov, A.O. Lyakhov, *Proc. Natl. Acad. Sci.* **106**, 17640 (2009).
23. R.T. Howie *et al.*, *Phys. Rev. B* **86**, 064108 (2012).
24. J.P.Perdew, K.Burke, M. Ernzerhof, *Phys. Rev. Lett.* **78**, 3865 (1996).
25. G. Kresse, J. Furthmuller, *Comp. Mater. Sci.* **6**, 15 (1996).
26. P.E. Blochl, *Phys. Rev. B* **50**, 17953 (1994).
27. W.Tang, E.Sanville, G.Henkelman, *J. Phys.: Condens. Matter.* **21**, 084204 (2009).
28. P. Giannozzi *et al.*, *J. Phys-Condens Mat.* **21,** 395502 (2009).
29. A.F. Goncharov *et al.*, *J. Synchrotron Rad.* **16,** 769 (2009).
30. A.F. Goncharov et al., *Physics Earth Planet. Interior* **174**, 24 (2009).
31. A.F.Goncharov, J. C. Crowhurst, V. V. Struzhkin, R. J. Hemley, *Phys. Rev. Lett.* **101**, 095502 (2008).
32. V. B.Prakapenka *et al.*, *High Pressure Res.* **28**, 225 (2008).

**Acknowledgments:** We thank the National Science Foundation (EAR-1114313) and DARPA (grant W31P4Q1210008) for financial support. A.F.G. acknowledges support from the NSF, Army Research Office, and EFRee- BES, EFRC center at Carnegie. Calculations were performed on the CFN cluster and Blue Gene supercomputer (Brookhaven National Laboratory, USA), on supercomputers of Moscow State University (Russia) and the Joint Supercomputer Center of Russian Academy of Sciences. Experiments were performed at GeoSoilEnviroCARS (Sector 13), Advanced Photon Source (APS), Argonne National Laboratory. GeoSoilEnviroCARS is supported by the National Science Foundation - Earth Sciences (EAR-1128799) and Department of Energy - Geosciences (DE-FG02-94ER14466)**.** Use of the Advanced Photon Source was supported by the U. S. Department of Energy, Office of Science, Office of Basic Energy Sciences, under Contract No. DE-AC02-06CH11357. A.R.O. designed the research. W.W.Z., Q.Z., S.E.B. and A.R.O. performed the calculations, interpreted data and wrote the paper. A.L. wrote the latest version of the structure prediction code. A.F.G., M.S., and V.P. did experiments. W.W.Z. and A.R.O contributed equally to this paper.

**Figure 1. Stability of new sodium chlorides:** (A) Pressure-composition phase diagram of the Na-Cl system. (B) Convex hull diagram for Na-Cl system at selected pressures. Solid circles represent stable structures; open circles metastable structures.

**Figure 2. Crystal structures of NaCl$_3$ and NaCl$_7$:** (A) *Pnma*-NaCl$_3$, (B) *Pm3*-NaCl$_7$, (C) *Pm3n*-NaCl$_3$. Large and small spheres represent Na and Cl atoms, respectively.

**Figure 3. Electronic structure of NaCl$_7$ and NaCl$_3$ at 200 GPa.** Band structure and density of states of (A) NaCl$_7$ and (B) NaCl$_3$, and electron localization function of (C) NaCl$_7$ and (D) NaCl$_3$. For clarity, atom-projected DOSs in (A, B) were multiplied by 3 (for NaCl$_7$) and by 4 (for NaCl$_3$).

**Figure 4. Crystal structures of:** (A) *P*4/*mmm*-Na$_3$Cl, (B) *P*4/*m*-Na$_3$Cl$_2$, (C) *Cmmm*-Na$_3$Cl$_2$, (D) *P*4/*mmm*-Na$_2$Cl, (E) *Cmmm*-Na$_2$Cl, and (F) *Imma*-Na$_2$Cl.

**Figure 5. Experimental data on NaCl$_3$ at 60 GPa. Powder X-ray diffraction pattern of NaCl$_3$ in Cl$_2$ medium at 60 GPa** (the X-ray wavelength is 0.5146 Å). Grey vertical ticks correspond to Bragg peak positions of Cl$_2$ and grey vertical ticks to those of NaCl$_3$.

**Table 1.** Structures of *Pnma*-NaCl$_3$ optimized at 40 GPa, A15-type NaCl$_3$ and NaCl$_7$, optimized at 200 GPa, and the corresponding atomic Bader charges (Q) and volumes (V).

A

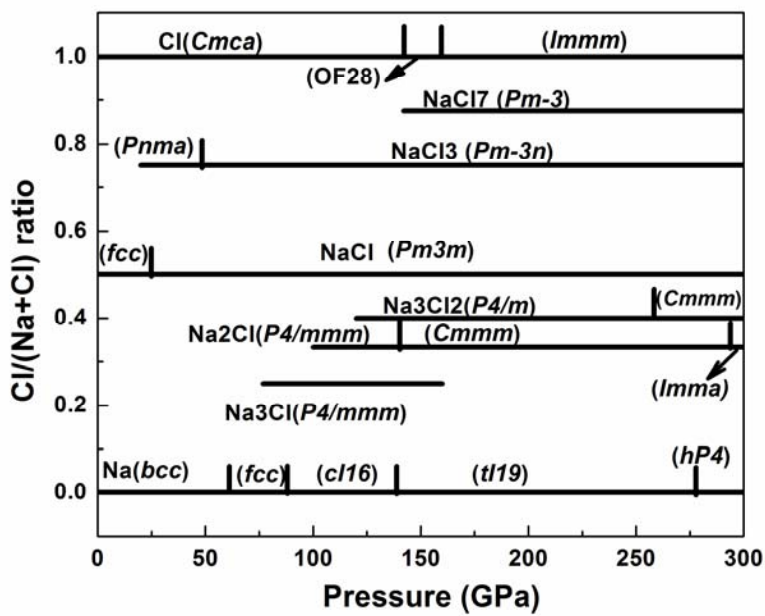

B

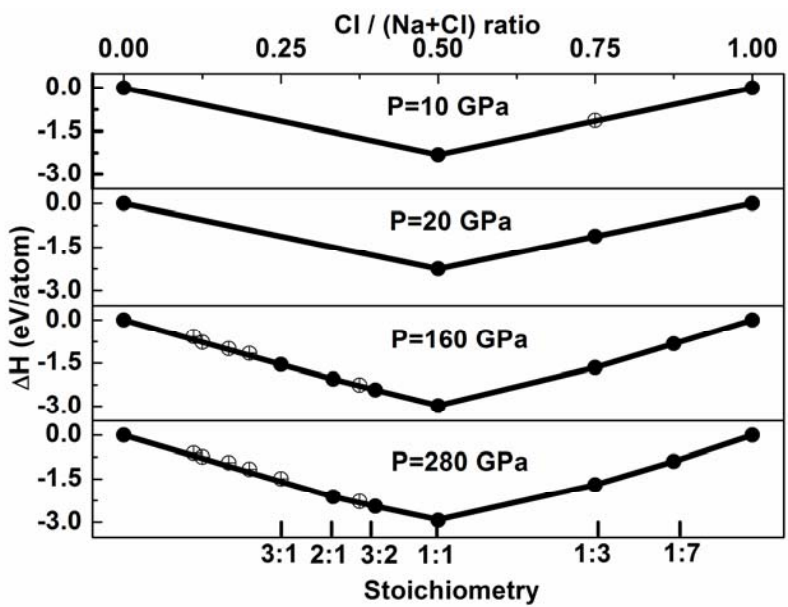

**Figure 1. Stability of new sodium chlorides:** (a) Pressure-composition phase diagram of the Na-Cl system. (b) Convex hull diagram for Na-Cl system at selected pressures. Solid circles represent stable compounds; open circles - metastable compounds.

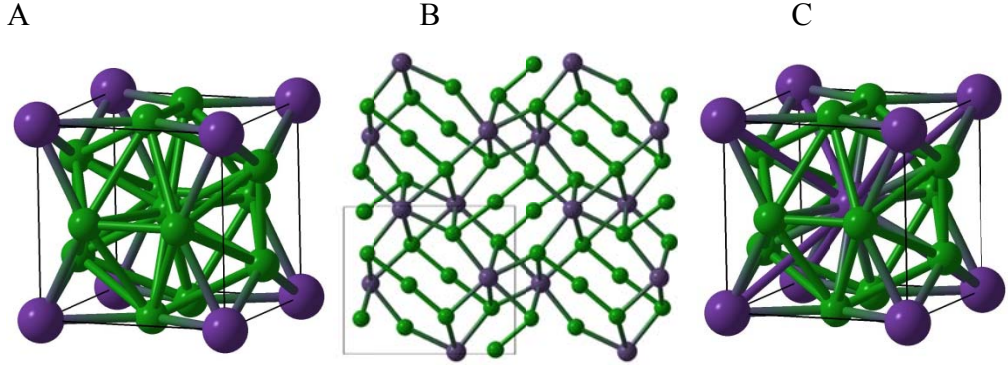

**Figure 2. Crystal structures of NaCl$_3$ and NaCl$_7$:** (A) *Pm3*-NaCl$_7$, (B) *Pnma*-NaCl$_3$, (C) *Pm3n*-NaCl$_3$. Blue and green spheres – Na and Cl atoms, respectively.

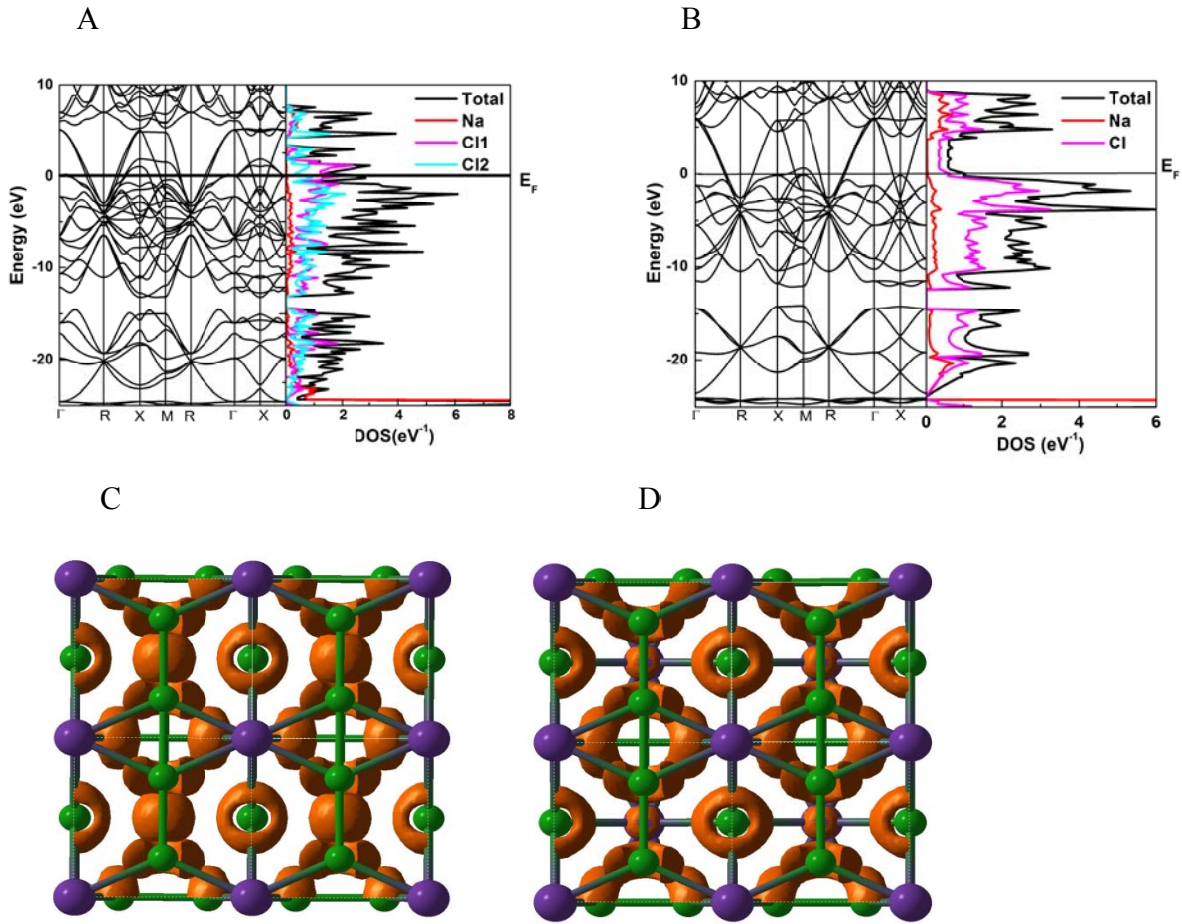

**Figure 3. Electronic structure of NaCl$_7$ and NaCl$_3$ at 200 GPa.** Band structure and density of states of (A) NaCl$_7$ and (B) NaCl$_3$ and electron localization function of (C) NaCl$_7$ and (D) NaCl$_3$. For clarity, atom-projected DOSs in (A, B) were multiplied by 3 (for NaCl$_7$) and by 4 (for NaCl$_3$).

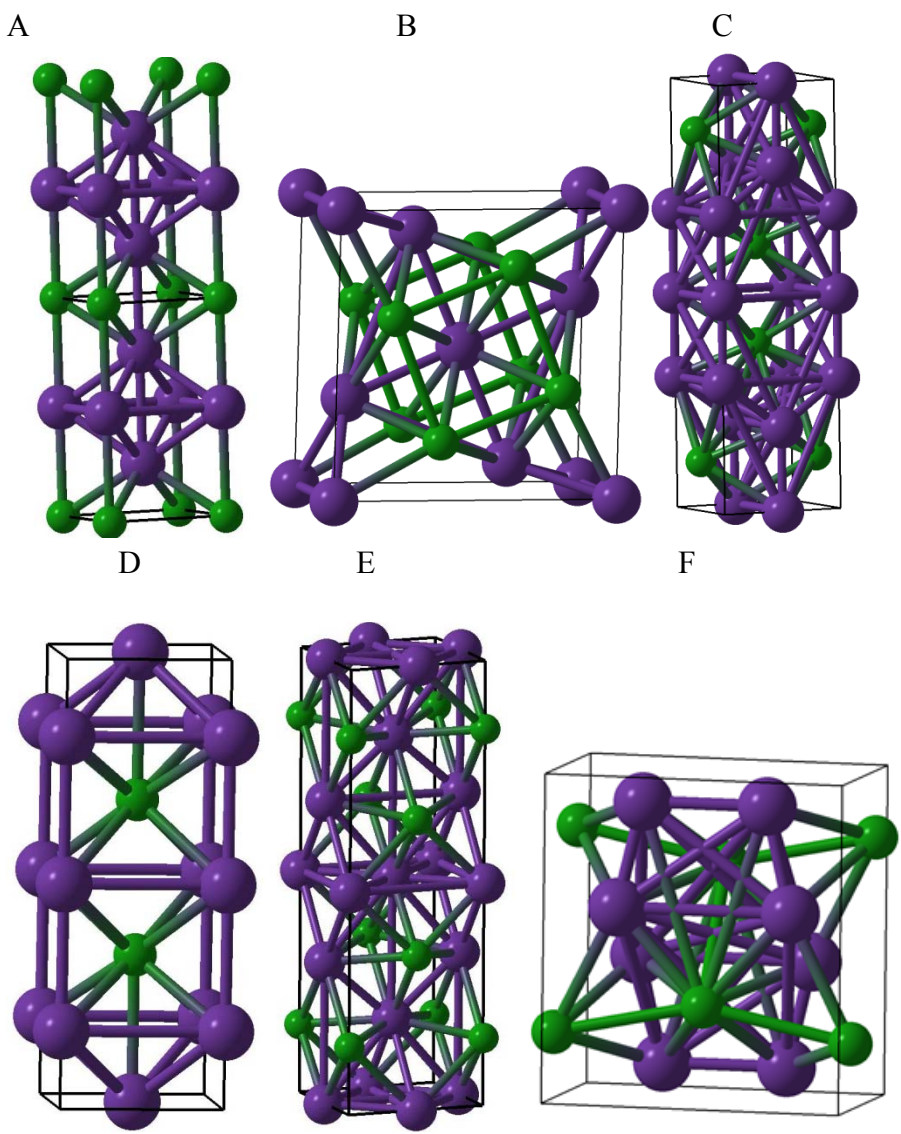

**Figure 4. Crystal structures of:** (A) $P4/mmm$-Na$_3$Cl, (B) $P4/m$-Na$_3$Cl$_2$, (C) $Cmmm$-Na$_3$Cl$_2$, (D) $P4/mmm$-Na$_2$Cl, (E) $Cmmm$-Na$_2$Cl, and (F) $Imma$-Na$_2$Cl.

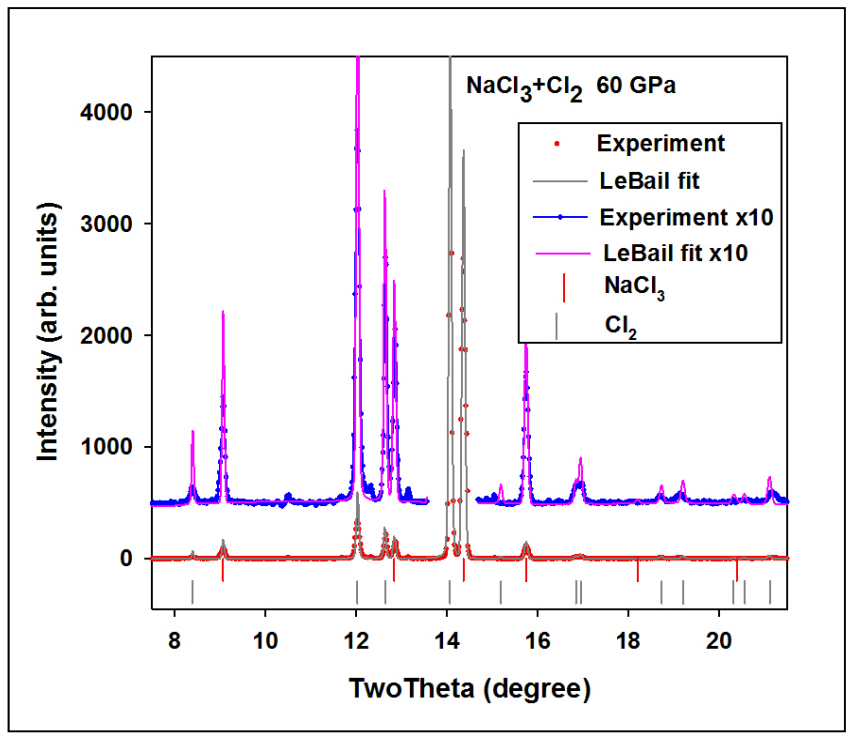

**Figure 5. Powder X-ray diffraction pattern of NaCl$_x$ in Cl$_2$ medium at 60 GPa** (the X-ray wavelength is 0.5146 Å). Grey vertical ticks correspond to Bragg peak positions of Cl$_2$ and grey vertical ticks to those of NaCl$_3$.

**Table 1.** Structures of *Pnma*-NaCl$_3$ at 40 GPa, and A15-type (*Pm3n*) NaCl$_3$ and NaCl$_7$ at 200 GPa, and the corresponding atomic Bader charges (Q) and volumes (V).

|  | Wyckoff position | x | y | Z | Q, |e| | V, Å$^3$ |
|---|---|---|---|---|---|---|
| NaCl$_7$ | Na(1a) | 0.000 | 0.000 | 0.000 | +0.827 | 4.30 |
|  | Cl(1b) | 0.500 | 0.500 | 0.500 | +0.067 | 8.63 |
|  | Cl(6g) | 0.248 | 0.500 | 0.000 | -0.149 | 9.61 |
| *Pm3n*-NaCl$_3$ | Na(2a) | 0.000 | 0.000 | 0.000 | +0.823 | 4.16 |
|  | Cl(6d) | 0.250 | 0.500 | 0.000 | -0.275 | 9.90 |
| *Pnma*-NaCl$_3$ | Na(4c) | 0.158 | 0.250 | 0.979 | +0.833 | 6.77 |
|  | Cl(4c) | 0.141 | 0.250 | 0.180 | -0.524 | 17.67 |
|  | Cl(4c) | 0.095 | 0.250 | 0.741 | -0.035 | 14.67 |
|  | Cl(4c) | 0.404 | 0.250 | 0.241 | -0.274 | 16.27 |

**Supplementary Materials:**

Methods
Supplementary Text
Figs.S1 to S7

# Supplementary Materials for

## Exotic thermodynamically stable sodium chlorides under pressure

Weiwei Zhang[1,2,*], Artem R. Oganov[2,3,*], Alexander F. Goncharov[4], Qiang Zhu[2], Salah Eddine Boulfelfel[2], Andriy O. Lyakhov[2], Maddury Somayazulu[4], Vitali Prakapenka[5]

*To whom correspondence should be addressed.

E-mail: zwwjennifer@gmail.com, artem.oganov@stonybrook.edu

[1]College of Science, China Agricultural University, Beijing, 100080, P.R.China. [2]Deptartment of Geosciences, Deptartment of Physics and Astronomy, State University of New York at Stony Brook, Stony Brook, NY 11794-2100, U.S.A. [3]Geology Department, Moscow State University, 119992 Moscow, Russia. [4]Geophysical Laboratory, Carnegie Institution of Washington, 5251 Broad Branch Road NW, Washington, D.C. 20015, U.S.A. [5]Center for Advanced Radiation Sources, The University of Chicago, Chicago, Illinois 60637, U.S.A. *These authors contributed equally to this work.

**This PDF file includes:**

**Methods**

**Supplementary Text**

**Figs.S1 to S7**

**Methods**

Structure/composition predictions were done using the USPEX code (*15-17*) in the variable-composition mode (*18*). The first generation of structures was produced randomly and the succeeding generations were obtained by applying heredity, atom transmutation, and lattice mutation operations, with probabilities of 60%, 10% and 30%, respectively. 80% non-identical structures of each generation with the lowest enthalpies were used to produce the next generation. All structures were relaxed using density functional theory (DFT) calculations within the Perdew-Burke-Ernzerhof (PBE) (*24*), as implemented in the VASP code (*25*). We used the all-electron projector augmented wave (PAW) (*26*) with [He] core (radius 1.45 a.u.) for Na and [Ne] core (radius 1.50 a.u.) for Cl, plane-wave basis sets with the 980 eV cutoff, and dense Monkhorst-Pack meshes with resolution $2\pi \times 0.05 \text{Å}^{-1}$. Having identified the most stable compositions and structures, we relaxed them at pressures from 1 atm to 300 GPa with an even denser Monkhorst-Pack mesh with resolution $2\pi \times 0.03 \text{Å}^{-1}$. Bader charge analysis was done using the grid-based algorithm (*27*) with 120x120x120 grids.

Phonons and Raman spectra were computed within density-functional perturbation theory as implemented in the Quantum Espresso package (*28*). We used a plane-wave kinetic energy cutoff of 180 Ry, in conjunction with dense k- and q-point meshes. For example, for *Pm3n*-NaCl$_3$, we used a 12×12×12 mesh for the Brillouin zone, and 4×4×4 *q*-mesh for computing the force constants matrix. The electron-phonon coupling matrix elements for the electron-phonon interaction coefficients were calculated with a large 16×16×16 grid.

Raman studies were performed using 488 and 532 nm lines of a solid-state laser. The laser probing spot dimension was 4 μm. Raman spectra were analyzed with a spectral resolution of 4 cm$^{-1}$ using a single-stage grating spectrograph equipped with a CCD array detector. Optical absorption spectra in visible and near IR spectral ranges were measured using an all-mirror custom microscope system coupled to a grating spectrometer equipped with a CCD detector (*29*). Laser heating was performed in a double-sided laser heating system combined with a confocal Raman probe (*30, 31*). Temperature was determined spectroradiometrically. Synchrotron XRD data were collected using bending magnet beamlines of GeoSoilEnviroCARS and of HPCAT at the Advanced Photon Source (*32*). The X-ray probing beam size was about 10 μm

**Supplementary text**

**Brief descriptions of the crystal structures.**

**NaCl$_7$** is stable above 142 GPa and has a cubic structure (space group *Pm*3) with 1 formula unit (f.u.) in the primitive cell. At 200 GPa it has the optimized lattice parameter *a* = 4.133 Å, with Na atoms occupying the Wyckoff 1a (0.0, 0.0, 0.0) position; there are two inequivalent Cl sites – Cl1 atoms occupy the 1b (0.5, 0.5, 0.5) and Cl2 the 6g (x, 0.5, 0.0) site with x=0.248. Na atoms sit at the corners of the cubic unit cell, while Cl2 atoms form an icosahedron around the Cl1 atoms. The nearest distance between Cl1 and Cl2 is 2.31 Å, whereas the shortest Cl2-Cl2 distance is 2.05 Å. This structure is shown in Fig. 2(b) and can be described as a derivative of the A15 (β-W or Cr$_3$Si) structure type.

**NaCl$_3$-*Pnma*** structure is stable at 25-48 GPa. It has 4 f.u. in the unit cell. This structure at 40 GPa has parameters *a* = 7.497 Å, *b* = 4.539 Å, *c* = 6.510 Å, with all atoms occupying Wyckoff 4c (x, 0.25, z) sites with x=0.158 and z= 0.979 for Na, x = 0.141 and z = 0.180 for Cl1, x = 0.095 and z = 0.741 for Cl2, and x = 0.404 and z = 0.241 for Cl3.

**NaCl$_3$–*Pm*3*n*** is predicted to become stable above 48 GPa. This is a metal with an A15 (Cr$_3$Si-type) structure with space group *Pm*3*n* and 2 f.u. in the primitive cell. At 200 GPa, the *Pm*3*n*-NaCl$_3$ structure has the optimized lattice parameter *a* = 4.114 Å with Na and Cl occupying the Wyckoff 2a (0.0, 0.0, 0.0) and 6d (0.25, 0.5, 0.0) positions, respectively. The nearest Na-Cl distance is 2.30 Å, whereas the shortest Cl-Cl distance is 2.06 Å - only slightly longer than the bond length in the Cl$_2$ molecule (1.99 Å). However, here and in NaCl$_7$, there are no molecules and short Cl-Cl bonds form extended monatomic chains running along the three mutually perpendicular axes.

***P*4/*mmm*-Na$_3$Cl** at 140 GPa has lattice parameters a= 2.786 Å and c=4.811 Å, with Na occupying the Wyckoff 2h (0.5, 0.5, 0.238) and 1b (0.0, 0.0, 0.5) positions, and Cl atoms at 1a (0.0, 0.0, 0.0) sites.

*Pm*3*m*-NaCl at 140 GPa has lattice parameter *a*=2.667 Å, and Na and Cl occupy the 1a (0.0, 0.0, 0.0) and 1b (0.5, 0.5, 0.5) sites, respectively.

***P*4/*m*-Na$_3$Cl$_2$** structure at 140 GPa has lattice parameters *a*=5.753 Å and *c*=2.820 Å, with Na atoms occupying the Wyckoff 4k (0.901, 0.676, 0.5), 1d (0.5, 0.5, 0.5), and 1a (0.0, 0.0, 0.0) positions, and Cl at 4j (0.799, 0.372, 0.0) sites.

***Cmmm*-Na$_3$Cl$_2$** structure at 280 GPa has lattice parameters *a*=9.881 Å, *b*=2.925 Å and *c*=2.508 Å with Na atoms occupying the Wyckoff 4g (x, 0, 0), 2b (0.5, 0.0, 0.0) positions with x= 0.196 and Cl 4h (x, 0.0, 0.5) at sites with x=0.115.

***P4/mmm*-Na$_2$Cl** at 120 GPa has lattice parameters *a*=2.790 Å and *c*=7.569 Å, with Na atoms occupying the Wyckoff 2h (0.5, 0.5, 0.827), 1d (0.5, 0.5, 0.5), and 1a (0.0, 0.0, 0.0) positions, and Cl sitting at 2h (0.0, 0.0, 0.673) sites.

***Cmmm*-Na$_2$Cl** at 180 GPa has lattice parameters *a*= 3.291 Å, b=10.385 Å, *c*=2.984 Å, and Na atoms at the Wyckoff 4j (0.5, 0.181, 0.5), 2d (0.5, 0.5, 0.5), and 2b (0.0, 0.5, 0.0) positions, and Cl in the 4i (0.0, 0.153, 0.0) sites.

***Imma*-Na$_2$Cl** at 300 GPa has lattice parameters *a*= 4.929 Å, *b*=3.106 Å, *c*=5.353 Å, with Na and Cl atoms occupying the Wyckoff 8i (0.791, 0.75, 0.588) and 4e (0.0, 0.25, 0.798) positions, respectively.

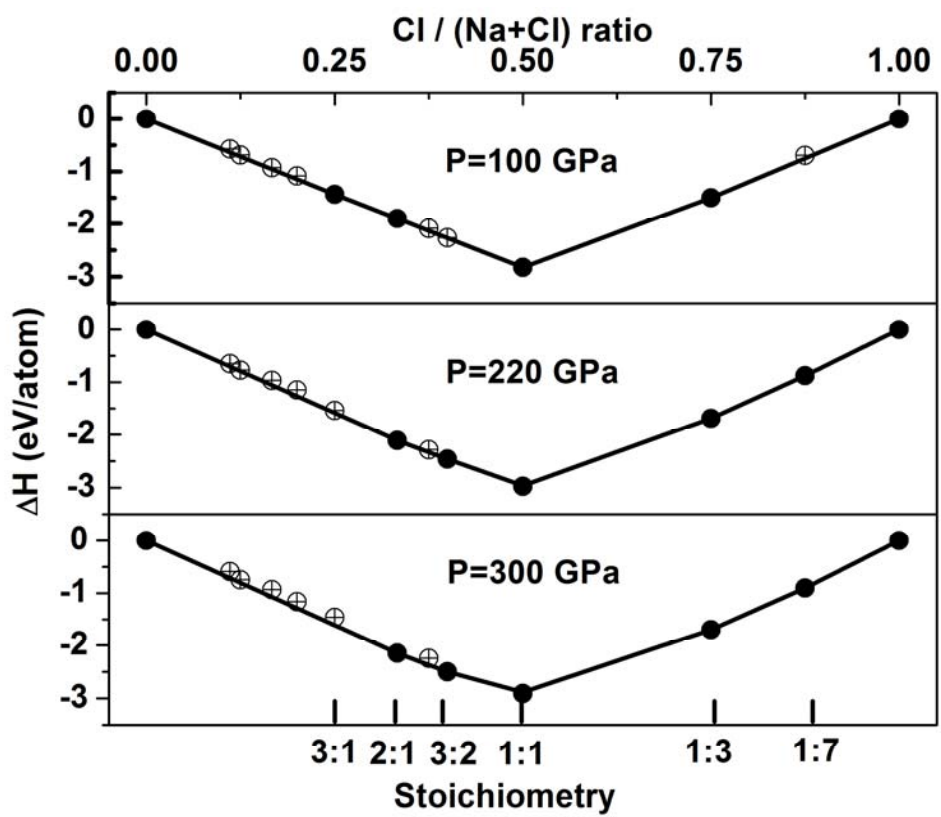

Figure S1. Convex hull diagrams for the Na-Cl system at several pressures. Solid circles represent stable structures; open circles metastable structures.

A

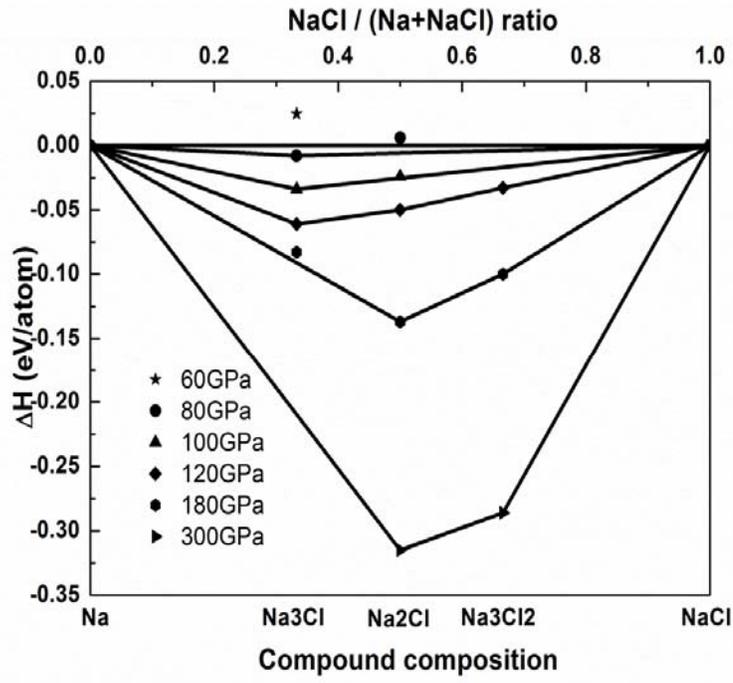

B

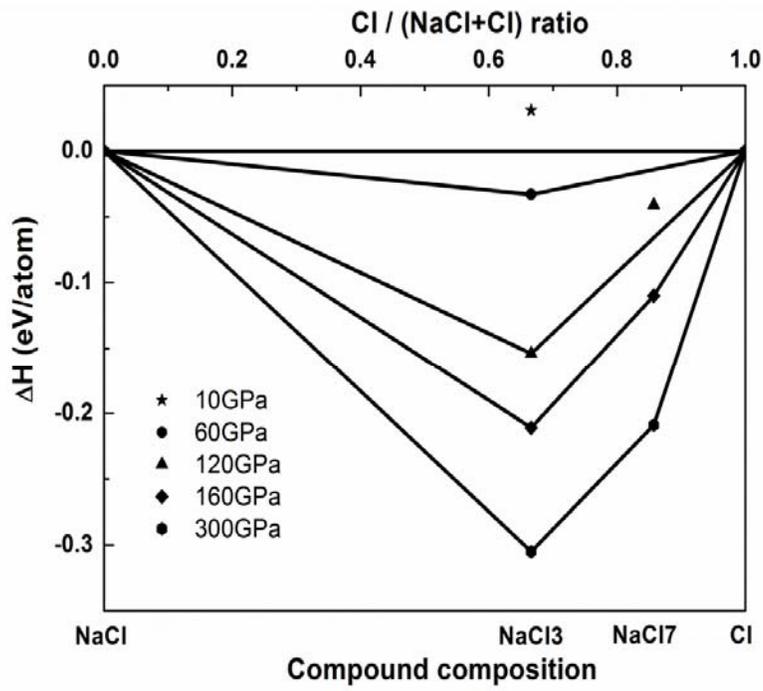

Figure S2. Convex hull diagrams for the (A) Na-NaCl and (B) NaCl-Cl systems. This illustrates that the formation of new sodium chlorides is a strongly exothermic process.

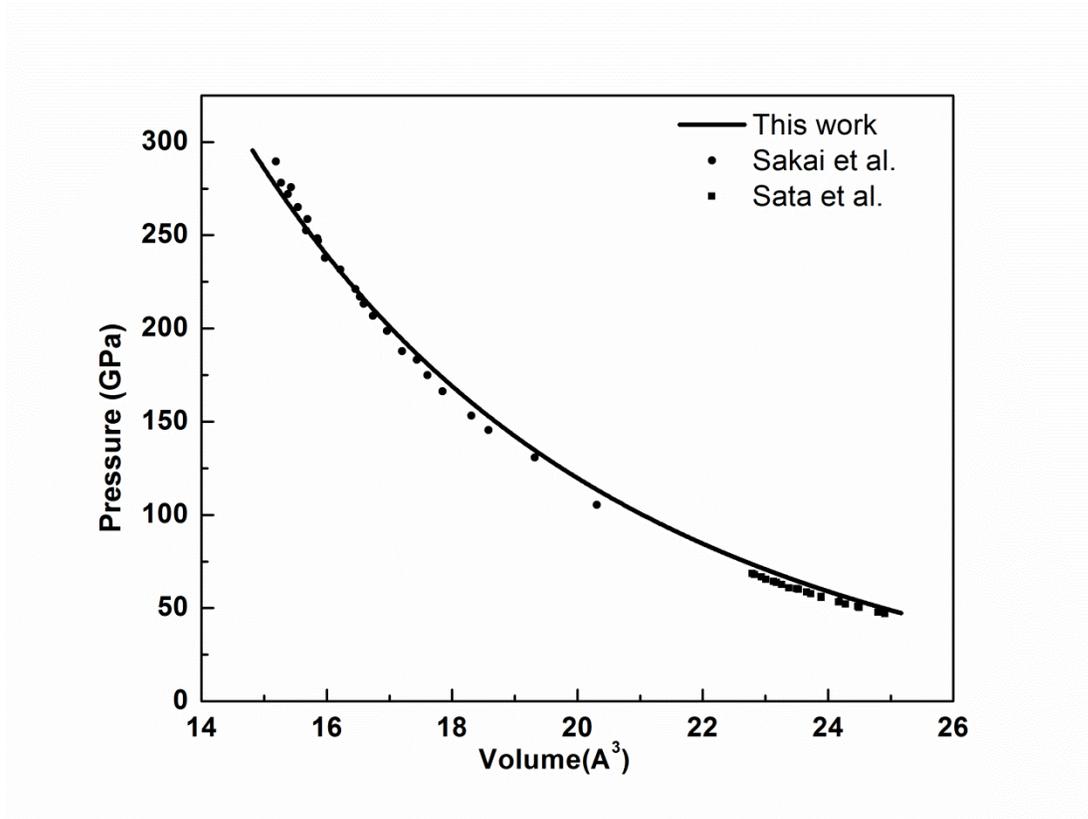

Figure S3. Equation of state of B2-NaCl from experiment (*4, 5*) and theory (this work). The solid line is a 3$^{rd}$-order Birch-Murnaghan fit to the energy-volume data from 40 GPa to 300 GPa. Circles are from experiments of Ref.5 with pressure determined using the equation of state of platinum. Squares are from experiments of Ref.4 with pressure determined using the equation of state of MgO. This simple comparison shows how accurately theoretical calculations model high-pressure behavior of NaCl.

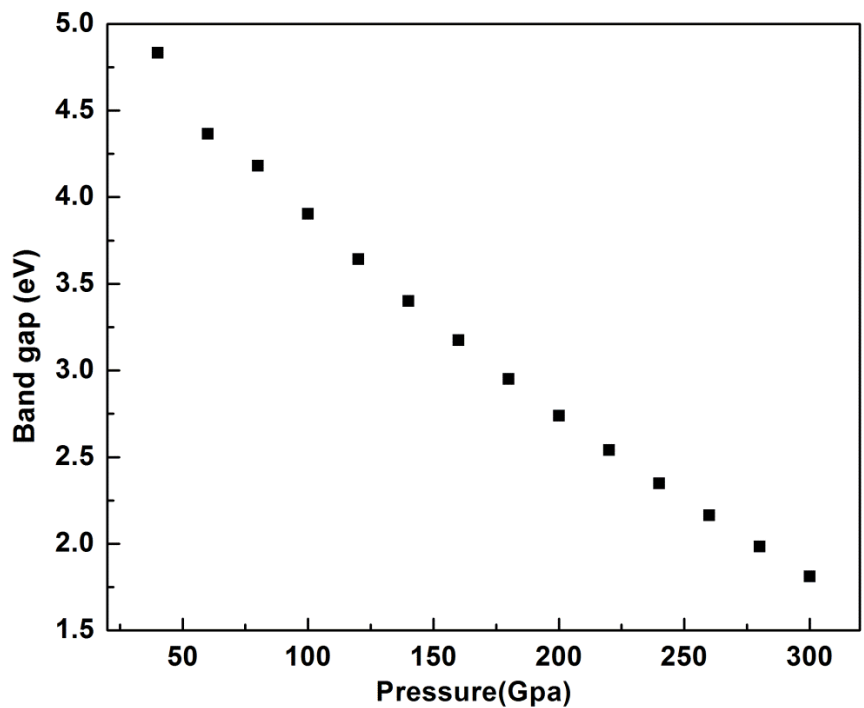

Figure S4. The M-M direct band gap of B2-NaCl as a function of pressure. This figure shows the computed DFT band gaps, which are lower bounds of the true gaps, and implies metallization of NaCl at multimegabar pressures.

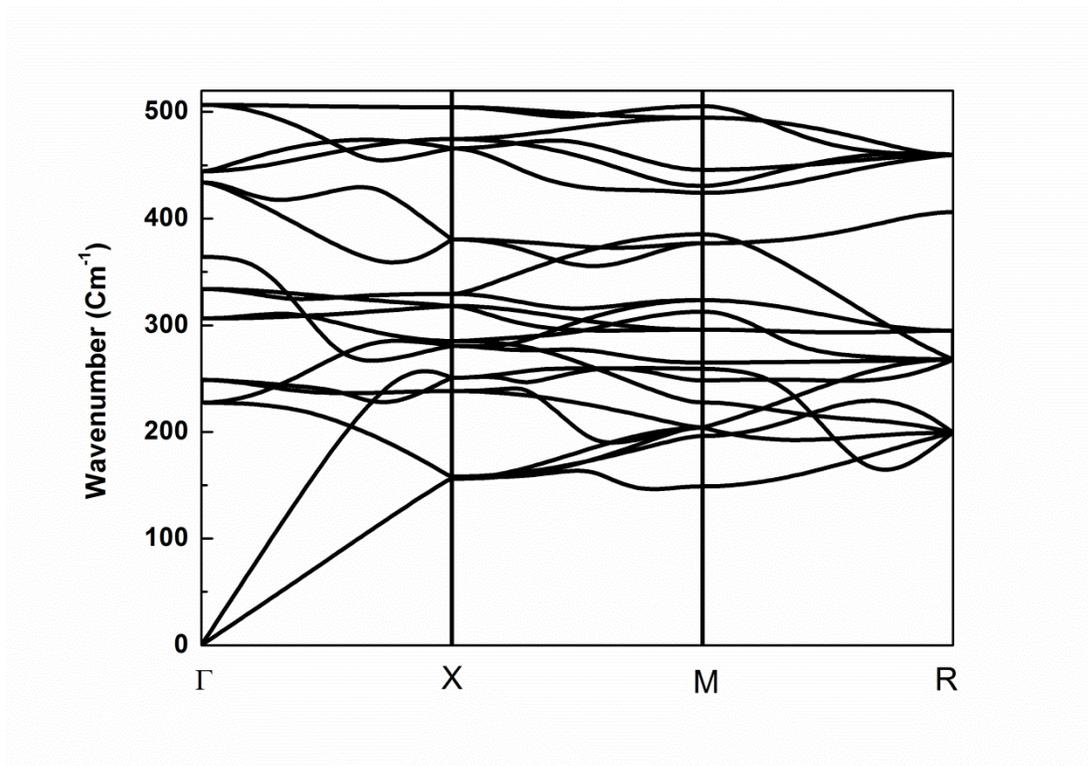

Figure S5. Phonon dispersion curves of cubic NaCl$_3$ at 60 GPa. Such calculations were done for all predicted structures to ensure their dynamical stability.

A                                    B

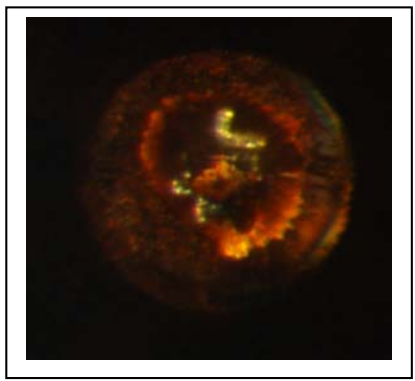

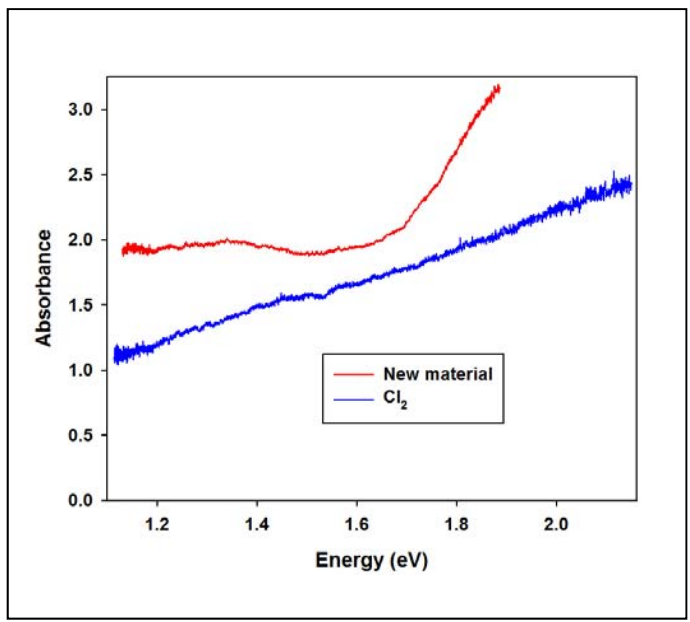

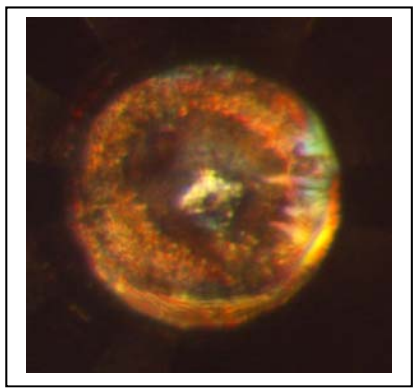

Figure S6. Optical properties of a material synthesized at 60 GPa and 2000 K in the DAC. (A) Optical photographs of the sample in transmitted and reflected light (top) and only reflected (bottom) light, demonstrating strong reflectivity. (B) Optical absorption spectra of a new material in comparison to that of $Cl_2$ medium. The former one can have a background contribution from unreacted chlorine medium. The latter we monitored as a function of pressure and showed a monotonous red shift without a substantial change in shape.

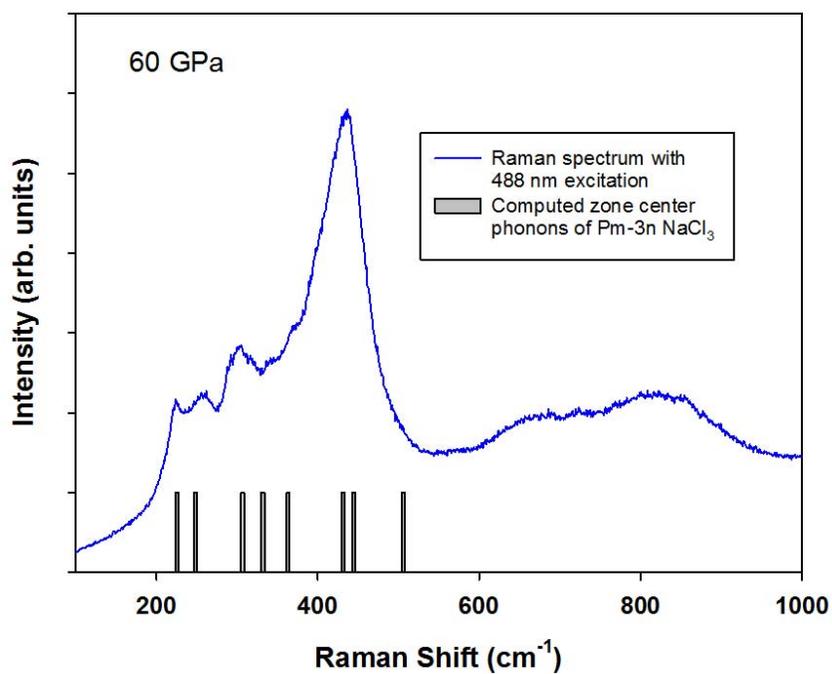

Figure S7. Raman spectrum on sample quenched to 300 K. Vertical bars correspond to the positions of the Brillouin-zone-center optical phonons of the cubic $Pm3n$-$NaCl_3$ computed in this work. Broad Raman peaks at 670 and 820 cm$^{-1}$ can be interpreted as second-order scattering (overtones and combination bands). Other probing spots show the presence of Raman peaks of pure $Cl_2$.